# Enhanced Mesenchymal Stem Cell Response with Preserved Biocompatibility via (MnZn)Ferrite–Polyacrylonitrile Composite Nanofiber Membranes

Baranサrac[a,b]*, Elham Sharifikolouei[c], Matej Micusik[d], Alessandro Scalia[c], Ziba Najmi[c], Andrea Cochis[c], Lia Rimondini[c], Gabriele Barrera[e], Marco Coisson[e], Selin Gümrükcü[f], Eray Yüce[a,g] and A. Sezai Sarac[h]

[a] Erich Schmid Institute of Materials Science, Austrian Academy of Sciences (ÖAW), 8700 Leoben, Austria

[b] Chair of Casting Research, Montanuniversität Leoben, 8700 Leoben, Austria

[c] Università del Piemonte Orientale UPO, Department of Health Sciences, Center for Translational Research on Autoimmune and Allergic Diseases-CAAD, 28100 Novara, Italy

[d] Polymer Institute, Slovak Academy of Sciences, Dubravska cesta 9, Bratislava 84541, Slovakia

[e] Istituto Nazionale di Ricerca Metrologica (INRiM), Strada delle Cacce 91, Torino, 10135 Italy

[f] Department of Chemistry, Istanbul Technical University, 34469 Istanbul, Türkiye

[g] Department of Materials Science, Chair of Materials Physics, Montanuniversität Leoben, 8700 Leoben, Austria

[h] Polymer Science & Technology, Istanbul Technical University, Maslak, 34469 Istanbul, Turkey

* E-mail: baransarac@gmail.com

**Abstract**

This study focuses on the synthesis and characterization of advanced polymeric composite electrospun nanofibers (NFs) containing magnetic oxide nanoparticles (NPs). By leveraging the method of electrospinning, the research aims to investigate polymer composites with enhanced interfacial properties, improved double-layer capacitance, and adequate biocompatibility. Electrospun polyacrylonitrile (PAN) NFs embedded with $Fe_2O_3$ and MnZn ferrite NPs were comprehensively characterized using advanced techniques, i.e., Fourier transform infrared spectroscopy, X-ray photoelectron spectroscopy, high-resolution scanning electron microscopy, X-ray diffraction, and alternating gradient field magnetometry. The incorporation of metal oxide NPs led to significant changes in the thermal, spectroscopic, and morphological properties of the NFs. XPS analysis confirmed increased oxidation, graphitic carbon content, and the formation of new nitrogen functionalities after heat treatment. Furthermore, interactions between nitrile groups and metal ions were observed, indicating the influence of nanoparticles on surface chemistry. Magnetic characterization demonstrated the potential of these composite NFs to generate magnetic fields for biomedical manipulation. Cytocompatibility studies revealed no significant impact on the viability or morphology of human mesenchymal stromal cells, highlighting their biocompatibility. These findings suggest the promising use of PAN-magnetic NFs in applications including targeted drug administration, magnetic resonance imaging, and magnetic hyperthermia for cancer treatment.

**Graphical Abstract**

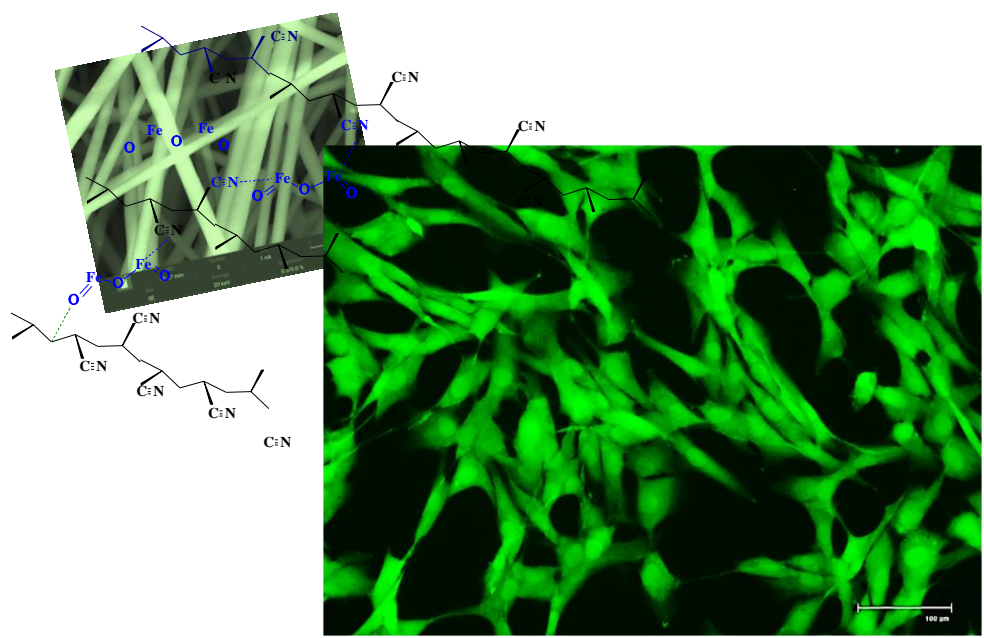

## 1. Introduction

Iron oxide in fibrous membranes can significantly influence human mesenchymal cell (hMSC) attachment and distribution [1, 2]. Iron oxide nanoparticles (IONPs) embedded in fibrous membranes can modify the surface roughness as well as mechanical, thermal and chemical properties of the membranes [3]. Iron oxide is biocompatible and can promote cell viability, where the inclusion of such nanoparticles into polymeric scaffolds can enhance cell adhesion by providing more binding sites for cell attachment. The presence of iron oxide can support cell attachment without causing cytotoxic effects, making it appropriate for biomedical uses [4-8]. The magnetic nature of iron oxide can be used to manipulate cell distribution and orientation using external magnetic fields. This property may be advantageous for applications in tissue engineering where ultimate control over cell placement is needed [9]. Iron oxide nanoparticles have been found to advance the osteogenesis of human mesenchymal stem cells (hMSCs). For regenerative medicine and tissue engineering applications, the attachment and subsequent differentiation of hMSCs are critical, making this particularly useful for bone tissue engineering [10]. Incorporating iron oxide nanoparticles can affect the nanostructure and porosity of the fibrous membranes. This can enhance nutrient and oxygen diffusion, promoting better cell growth and distribution [11].

Iron oxide embedded in polymeric fibers can promote angiogenesis, the process through which new blood vessels develop from existing ones. Iron oxide nanoparticles (IONPs) possess magnetic properties that can be used to enhance angiogenesis. The use of an external magnetic field guides and concentrates therapeutic agents, growth factors, or even cells to specific sites, thereby promoting localized angiogenesis [12, 13]. IONPs can be functionalized to carry and release angiogenic growth factors such as VEGF (vascular endothelial growth factor). When incorporated into polymeric fibers, these nanoparticles can enable a sustained release of growth factors, thereby promoting angiogenesis over an extended period. Iron oxide can improve the overall biocompatibility of the scaffold, enhancing cell viability and proliferation. This supportive environment can facilitate endothelial cell function, which is crucial for new blood vessel formation [14].

Polymeric nanofibers (NFs) have received a great deal of attention due to their remarkable attributes, i.e., well-defined molecular alignment, high porosity, large specific surface area, and unique nanoscale properties [15]. The functionality of electrospun fibers with adjustable polymer composite chemistry offers a versatile platform for exploring diverse applications, including filtration media, sound isolation materials, and sensor components [16, 17]. Nanofibers enable the easy integration of a wide range of therapeutic molecules, improving the capacity for drug loading and promoting controlled, long-term release. Additionally, by controlling the arrangement of polymeric fibers at a larger scale, new possibilities emerge for using them as implantable systems for localized drug delivery. [18]. Smart polymer nanofibers demonstrate rapid response times due to their unique structural properties. These characteristics make them ideal for utilization in a variety of applications, including serving as a targeted, implantable platform for controlled biomolecule and drug delivery [19-22]. Magnetic metal oxide nanoparticles can be mixed with the polymer matrix to promote electrical conductivity and thermal properties [23].

Several studies report that electrospun NFs—with diameters ranging from 50 to 1,000 nanometers and composed of materials such as collagen, laminin, polycaprolactone-chitosan blends, and polyvinyl alcohol–chondroitin sulfate—enhance cell viability and enable diverse responses including osteogenic, chondrogenic, and neural differentiation [24-28]. Another advantage is that, unlike traditional culture substrates, NFs are capable of advanced cell generation, such as spheroid formation and long-distance communication [29]. On the downside, until now, many studies have not fully reported scaffold properties and cell culture parameters, limiting reproducibility and cross-study comparison. Another disadvantage is the lack of long-term or in-vivo validation tests in many studies. Also, achieving the full biomimicry of the native extracellular matrix, particularly regarding hierarchical structure and dynamic signaling, remains a challenge [30]. Other factors such as scalability, cost and regulatory approvals impose a barrier for translation to clinical trials [26].

Polyacrylonitrile (PAN) is an important biocompatible polymer recognized for its non-toxicity, nonantigenic activity and ease of use in the development of bio-related nanomaterials. Its suitability for creating these materials is further enhanced by its ability to support the structural and temporal control of peptide self-assembly processes [31]. PAN is a polymeric material commonly employed in membrane technology because of its excellent chemical resistance and strong performance in water filtration applications [32]. Electrospun NF networks can be readily fabricated using PAN polymers. These resulting NF membranes feature a remarkably high specific surface area, a highly porous structure, and finely controlled pore sizes [33-35]. PAN is a polymer with partial crystallinity and notable polarity, commonly processed from solution-based methods. Computational analyses indicate that the monomer unit of PAN interacts with each solvent through dipole-dipole forces, leading to the formation of PAN–solvent complexes. In these complexes, the C≡N group of PAN aligns in an antiparallel orientation with the polar group of the solvent molecule (S═O or C═O group) [36]. Key characteristics such as the exceptionally high rate of polymerization in water, the polymer's ability to dissolve in highly concentrated solutions of inorganic salts, its high melting point, the significant reduction of the glass transition and melting temperatures by water, and its plasticization by polar ingredients are all linked to their molecular foundations. These phenomena are primarily attributed to the high noncovalent intra- and intermolecular interactions driven by the strong polar nitrile group. Furthermore, the contributions of dipole-dipole interactions, the formation of electron-donor-acceptor complexes, and hydrogen bonding have been investigated [37].

Production of conductive fibers blending with the metal oxide nanoparticle containing polyacrylonitrile-based composite fiber mats offers exciting potential for use in smart textiles, particularly in health and medical fields, as well as in energy applications, i.e., supercapacitors and batteries [38, 39]. Synthesizing and characterizing well-dispersed magnetic metal oxides at the nanometer scale has opened interesting possibilities in composite materials research [40]. Their specific electronic structure showcases unique stability and physical and chemical properties different from microparticles and bulk materials. When integrated into polymers that interact with environmental molecules and sunlight, they exhibit enhanced (photo)catalytic activity, electronic properties, and optical behavior [41]. Electrospun composite fibers have been explored for wastewater treatment. A branched polyethyleneimine (b-PEI) functionalized magnetic iron oxide ($Fe_3O_4$)/PAN electrospun fiber (b-PEI-FePAN) was developed to remove hexavalent chromium [Cr(VI)], with b-PEI grafted PAN serving as a flexible base and enhancing adsorption capacity [42]. Iron-loaded aminated polyacrylonitrile fiber (PANAF-Fe) was studied through a chemical grafting process for its ability to remove phosphate from wastewater [43]. Iron oxide nanoparticles are key in forming electrical networks, enabling fast electron transfer in anode materials. STM measurements show that hydrogen binds to the oxygen surface at unblocked sites, creating a hydrogen bond with an oxygen atom that is symmetrically identical [44].

The characteristics of these nanocomposites are significantly affected by the quantity, size, and distribution of the metal nanoparticles [45]. The PAN/Ag composite NFs exhibited greater breaking stress and extension before rupture relative to the pure PAN NFs. Additionally, the inclusion of silver nanoparticles enhanced their conductivity. Integrating magnetic metal nanoparticles (MGMNPs) with the constituent elements of nanocomposites within biocompatible polyacrylonitrile will enhance electron transport efficiency, where we aim to understand the molecular interaction of MGMNPs with the polyacrylonitrile structure by advanced characterization methods. Investigations of blood-surface interaction with polyacrylonitrile-based membranes during hemodialysis are reported [46]. Electrospun magnetic NF mats are utilized in magnetic hyperthermia therapy, enabling targeted drug delivery and serving as valuable diagnostic and therapeutic tools in cancer treatment. These mats offer innovative solutions for improving the precision and effectiveness of cancer therapies. [47].

PAN and Fe(III) metal-organic framework nanocomposites were created via electrospinning, enhancing cellular interactions with the substrate [48]. In medical magnetic particle imaging (MPI), magnetic nanoparticle contrast agents are safer than other materials with high contrast and better sensitivity [49]. Among them, iron oxide nanoparticle tracers are preferred for MPI. Magnetic nanoparticles have been extensively researched for their possible application in biomedical applications, particularly in hyperthermia for cancer treatment [50]. To maximize their effectiveness in this context, it is crucial to develop magnetic nanoparticles that offer high heating efficiency, as this enhances their ability to influence the intracellular environment and generate heat effectively. Thus, the design of magnetic nanoparticle composite NFs might have applications in imaging and magnetic field manipulation.

The primary objective of this study is to demonstrate that ferrite-containing PAN NF composites exhibit improved hMSC adhesion and proliferation compared to plain PAN NFs without compromising biocompatibility. This enhancement is attributed to the interaction between the PAN and magnetic NPs, thereby boosting double-layer capacitance, charge collection efficiency, and overall energy storage performance. PAN-magnetic iron oxide porous NF mats, with the aid of their porous structure, hold promise for both diagnostic and therapeutic applications, offering promising opportunities in various biomedical fields. Their compelling biocompatibility, mechanical robustness, and porous architecture further elevate their suitability for magnetic diagnostics. This enhanced suitability stems from the ability of the homogenously dispersed magnetic particles to effectively detect and monitor various medical conditions [51]. Additionally, iron oxide/poly(m-anthranilic acid)/poly(ε-caprolactone) (PCL) composite NF production and advanced characterizations [52], electrospun polyacrylonitrile/2-(acryloyloxy)ethyl Ferrocenecarboxylate (FcP) NFs [53], and the thermomechanical characteristics of magnetic nanoparticles confined within PAN NFs when subjected to a magnetic field [54] have been recently studied. In this study, electrospun NF containing metal oxide nanoparticles ($Fe_2O_3$ and MnZn ferrite), fabricated to achieve electrical conductivity and magnetic properties, have undergone comprehensive and advanced characterizations mainly via X-ray photoelectron spectroscopy (XPS), X-ray diffraction (XRD), Fourier transform infrared (FTIR) spectroscopy, and high-resolution scanning electron microscopy (SEM), for the comprehensive understanding the mechanism of interaction between metal oxide and nitrile group of polyacrylonitrile matrix reflecting to the influence of nanoparticles on surface chemistry. Magnetic characterization demonstrated the potential of these composite NFs to generate magnetic fields for biomedical manipulation. Cytocompatibility studies confirmed the materials' biocompatibility, as they did not significantly affect human mesenchymal stromal cell viability or morphology. Cell adhesion and proliferation tests evidenced the enhanced cell-biomaterial interaction with the use of composite NFs.

## 2. Experimental Section
### 2.1. Material

N,N'-dimethylformamide (DMF≥99.8%), polyacrylonitrile (PAN, $M_{w\_average}$ = 150,000), and indium tin oxide coated polyethylene terephthalate (ITO-PET) sourced from Sigma Aldrich were utilized in the experiments. Iron Oxide Nanopowder (γ-$Fe_2O_3$) and MnZnFerrite nanoparticules were sourced from US Research Nanomaterials, Inc. and were of analytical grade, utilized in their original form without any further purification.

### 2.2. Formation of PAN/Metal Oxide Nanoparticle Composite Nanofibers

The electrospinning precursor was formulated by completely dissolving PAN powder in a DMF-based solvent using a cup horn sonicator to create 100 mL of a 10% PAN solution. For the composite NFs, MnZn Ferrite (5 wt%, 99.99% purity, ~28 nm diameter near-spherical) and γ-$Fe_2O_3$ (5 wt%, 99.9% purity, ~10 nm diameter near-spherical) NPs were dispersed in this 10 ml of 10% PAN solution in the same cup horn sonicator by overnight stirring (12 h) to avoid sediment formation and to facilitate homogenous dispersion of the metal oxide NPs. The electrospinning setup consists of a grounded collector, a syringe pump with a flow rate of 5.5 µL $h^{-1}$ to 20 mL $h^{-1}$, and a high-voltage DC power supply capable of up to 50 kV. The 0.7 mm syringe needle had a positive electrode. The syringe pump precisely controlled the flow rate of the polymer solution, ensuring a steady supply during electrospinning. The polymer solutions were electrospun in a horizontal configuration onto an aluminum collector. To fabricate NFs, the electrospinning was conducted at room temperature with applied voltages between 10 and 15 kV. The capillary tip was consistently maintained at a distance of approximately 15 cm from the collector, while the solution was supplied at a rate of 1 mL $h^{-1}$. No visual sedimentation was observed in the syringe after each electrospinning trial, proving the even dispersion of the PAN and metal oxide NPs within the solvent. Heat treatment of some of the NFs was conducted inside a dynamic mechanical analyzer, as previously explained in [54]. In short, the specimens were subjected to a controlled temperature increase from 300 K to 800 K at a uniform 5 K $min^{-1}$ heating rate. The tests were performed under an Ar flow of 20 mL $min^{-1}$ and cooled down in a furnace environment after reaching the maximum temperature.

### 2.3. Electrochemical Impedance Spectroscopy

The standard three-electrode system was employed, consisting of PAN/Metal Nanoparticles NF mats as the working electrode, Ag/AgCl (3 M KCl) as an aqueous reference electrode, and platinum spiral wire as a counter electrode. Electrochemical impedance spectroscopy measurements were implemented over the frequency range of 100 kHz to 0.01 Hz via Gamry Reference 600 (USA). The EIS test was conducted under open-circuit potential (OCP) conditions. Two electrolyte mediums were used in the electrochemical measurements as 7.4 pH PBS buffer and 0.5 M $H_2SO_4$.

### 2.4. Structural Analysis

XRD measurement was conducted utilizing a Bruker D2Phaser diffractometer equipped with an LYNXEYE-2 detector, operating in Bragg-Brentano (θ-2θ) configuration. The measurements were performed over an angular range spanning from 5° to 100°, employing Co Kα radiation with a wavelength of 0.17902 nm. Data acquisition was carried out with a step size of 0.005°.

### 2.5. Spectroscopic Analysis

XPS measurements were conducted using a Thermo Scientific Nexsa G2 Surface Analysis System (Thermo Fisher Scientific, UK), featuring a micro-focused, monochromatic Al Kα X-ray source (1486.68 eV). The system utilized an X-ray beam with a diameter of 400 mm. The spectra were collected using a constant analyzer energy mode, employing a pass energy of 200 eV for the survey scans. For detailed measurements, smaller regions were analyzed using a 50 eV pass energy. Charge compensation during the analysis was accomplished through the system's dual beam flood gun. Data acquisition and processing were completed through Thermo Scientific Avantage software, version 6.9.0. Spectral calibration was performed using the automated calibration routine and internal standards of Au, Ag, and Cu provided by the K-Alpha system. The atomic percent surface compositions were assessed by analyzing the integrated peak areas of the identified elements and applying the corresponding sensitivity factors. The concentration fraction of element A was determined by:

$$\% A = \frac{I_A/s_A}{\sum(I_n/s_n)} \times 100\% \qquad , \qquad (1)$$

where $I_n$ and $s_n$ represent the total integrated peak areas and the Scofield sensitivity factors, both adjusted for the transmission efficiency of the analyzer, respectively. Attenuated total reflectance Fourier-transform infrared (ATR-FTIR) spectroscopy was carried out using a Bruker Vertex 70 ATR spectrometer, operating at a spectral resolution of 0.4 cm$^{-1}$ and spanning a wavenumber range of 600 to 4000 cm$^{-1}$.

### 2.6. Morphological Analysis

The morphological characteristics of the NFs that were produced were scrutinized by scanning electron microscopy. The investigation by energy dispersive X-ray (EDX) was conducted using a Quanta FEG 250 microscope equipped with an integrated EDX detector. SEM micrographs were captured at magnification levels of 20k×, 50k×, and 100k×. To ensure sufficient collection of Fe signals, an acquisition time of 20 minutes was set for the EDX mapping process.

### 2.7. Magnetic Analysis

The room-temperature magnetic properties were tested by a high-sensitivity alternating gradient field magnetometer (AGFM) (Princeton Measurements Corporation, Princeton, NJ, USA). Hysteresis loops were recorded within a ±10 kOe magnetic field range. The diamagnetic contribution of the sample holder was carefully accounted for and subtracted from the measured data.

### 2.8. Cytocompatibility Analysis

Human bone marrow-derived mesenchymal stromal cells (hBMSCs) were acquired from the American Type Culture Collection (ATCC PCS-500-012, $10^6$ viable cells per mL) and cultured in low-glucose DMEM (Sigma-Aldrich), enriched with 15% fetal bovine serum (FBS) and 1% antibiotics in an environment containing 5% $CO_2$ in a 24-well plate. When they attained 80–90% confluence, they were isolated with a trypsin-EDTA solution, harvested, and prepared. A specific number of cells ($1.5 \times 10^4$) was evenly distributed onto the material surfaces (1 cm × 0.5 cm nanofiber mats) and incubated for a few hours to ensure proper adhesion and spreading. Subsequently, the samples were submerged in a complete culture medium and returned to the incubator for 24 and 48 hours. At all relevant intervals, the cellular metabolic activity was assessed using the Alamar Blue assay, following the manufacturer's instructions. Metabolically active cells reduce the blue, nonfluorescent molecule resazurin into the red/violet fluorescent molecule resorufin in direct proportion to cell number. A 0.015% Alamar solution in a complete medium was applied to the seeded scaffolds and incubated for 3 hours in the dark. After incubation, 100 µl of the solution was moved to a black plate, and fluorescence was recorded by the Spark spectrophotometer (Tecan, Männedorf, Switzerland) through excitation and emission wavelengths of 530 nm and 590 nm, respectively. During the last time point, the viability of cells is further confirmed using the lactate dehydrogenase (LDH) assay (CyQUANT LDH Cytotoxicity Assay, fluorescence). LDH release was used as a marker of cytotoxicity. LDH activity in the culture medium was quantified via a coupled enzymatic reaction in which LDH catalyzes the conversion of lactate to pyruvate with concomitant $NAD^+$ reduction, followed by diaphorase-mediated resazurin reduction to fluorescent resorufin (Ex/Em: 560/590 nm). The signal is proportional to LDH levels. The test was conducted according to the manufacturer's instructions, and results are expressed as absolute LDH activity. Viable cells are finally visualized on the materials using the exclusive LIVE/DEAD® assay. This two-color fluorescence-based assay employs Calcein AM (green) and ethidium homodimer-1 (EthD-1, red). Live cells, characterized by high intracellular esterase activity, enzymatically convert the cell-permeant, nonfluorescent Calcein AM into highly fluorescent, green-emitting calcein (Ex/Em 495 nm/515 nm). Conversely, EthD-1 penetrates cells with compromised membranes and emits bright red fluorescence (Ex/Em 495 nm/635 nm). The intact plasma membranes of live cells exclude EthD-1, preventing red fluorescence. The EVOS FLoid Imaging System was employed to capture the images (Thermo Fisher, Massachusetts, USA).

**2.9. Cell morphology, adhesion, and proliferation assessment**

The morphology and distribution of cells in contact with the materials were evaluated using SEM. After the last time point, cells cultivated on top of the materials were fixed overnight (4°C) using a solution of glutaraldehyde (2.5% in PBS) to preserve their ultrastructure. Later, specimens were dehydrated using an increasing ethanol scale (70, 90, 100%-I, 100%-II for 1h each) to remove water from the cells progressively. Finally, specimens were immersed in hexamethyldisilazane (HMDS), a low-tension reagent that quickly dries the specimens, reducing the structural artifacts. Specimens were then coated with a thin layer of gold (10nm) using a Smart Coater (Jeol, Akishima, Japan) and observed using the JSM-IT500 InTouchScope™ SEM (Jeol).

Cell adhesion and proliferation of cells in contact with the materials were further investigated using RT-PCR and the Click-iT™ EdU proliferation assay (Invitrogen, Thermo Fisher), respectively. For gene expression analysis, total RNA was extracted, and the expression levels of Cadherin-1 (CDH1) and Ki67 were quantified to assess the scaffolds' ability to support hBMSC adhesion and proliferation, respectively. Together with Ki67, the hMSC proliferation was evaluated using the Click-iT™ EdU Proliferation Assay, which utilizes 5-ethynyl-2'-deoxyuridine (EdU), a nucleoside analog incorporated into newly synthesized DNA. Briefly, 24 hours after cell seeding onto the material surfaces, a 2 µM EdU solution was added and incubated overnight. EdU detection was achieved via a copper-catalyzed click reaction between the alkyne group of EdU and azide-modified horseradish peroxidase (HRP), followed by the addition of the Amplex™ UltraRed reagent. The resulting fluorescent signal was measured using the spark spectrophotometer (Ex/Em: 568/585 nm). Lastly, the maintenance of the stemness of cells cultivated on the materials was assessed by evaluating the expression of the stemness markers CD90 and CD105 via RT-PCR. Results and primer sequences are presented in Figure S2 and Supplementary Table S2. For all the RT-PCR analyses, the glyceraldehyde 3-phosphate dehydrogenase (GAPDH) was used as the housekeeping gene.

### 3. Results & Discussions
#### 3.1. Possible Interaction Mechanism

A long-term dispersion ensures that the iron oxide nanoparticles are evenly distributed within the PAN matrix. During this step, the chemical interactions (hydrogen bonding, electrostatic interactions, and coordination bonding) start to play a role in stabilizing the dispersion (Figure 1). The bonding electrons in the C–N bond are more attracted towards the N atom because of the higher electronegativity of the N atom. Hence, the bonding electrons concentrate more on the N atom, with a partial negative charge on the nitrogen in comparison to the partial positive charge on the carbon atom. The cyanide ion interacts with transition metals to create M–CN bonds (where M refers to Fe or MnZn). The strong attraction of metals to this anion can be explained by its partial negative charge, its small size, and its capacity to participate in π-bonding. This bonding occurs as the metal ion's d orbitals overlap with the π orbitals of the polar C≡N bond.

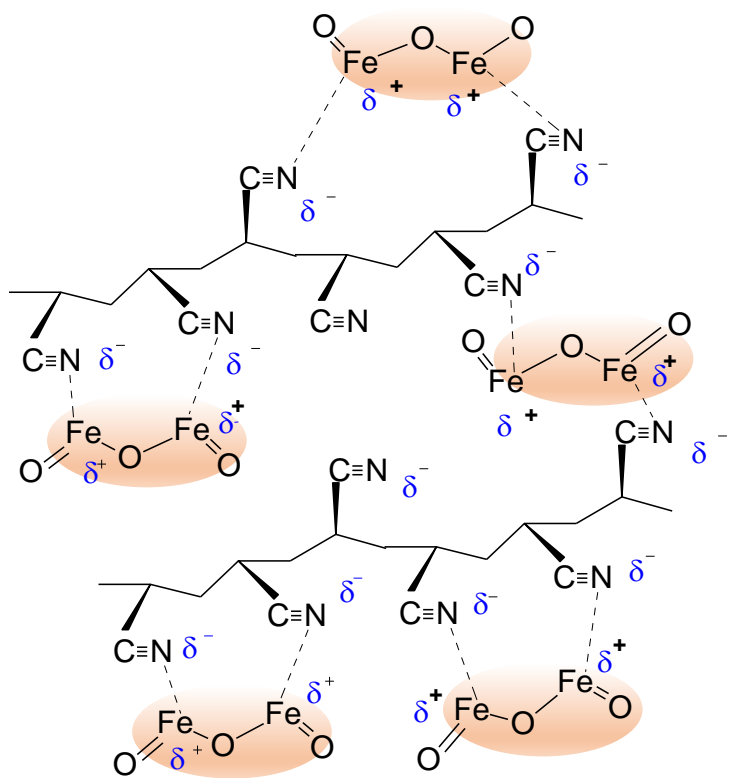

**Figure 1** – Scheme of possible interaction of $Fe_2O_3$ with Polyacrylonitrile

### 3.2. Morphology and Microstructure

High-resolution SEM images, particularly at high magnification, demonstrate the formation of uniform NFs along with the distribution of single nanoparticles. The PAN NFs exhibit a narrow size distribution, with diameters averaging 970±110 nm (Figure 2a). A homogeneous dispersion of NFs is observed over extensive areas of interest (Figure 2d). Fiber diameter size diminished when particles were included (520±50 nm and 520±90 nm for PAN/γ-$Fe_2O_3$ and PAN/MnZn-Ferrite, respectively). Partial clustering was observed for both γ-$Fe_2O_3$ (Figure 2b,e) and MnZn-Ferrite (Figure 2c,f) nanoparticles within PAN NFs, which is mainly because ceramic nanoparticles typically have high surface energies, leading to strong interparticle interactions that favor agglomeration overdispersion.

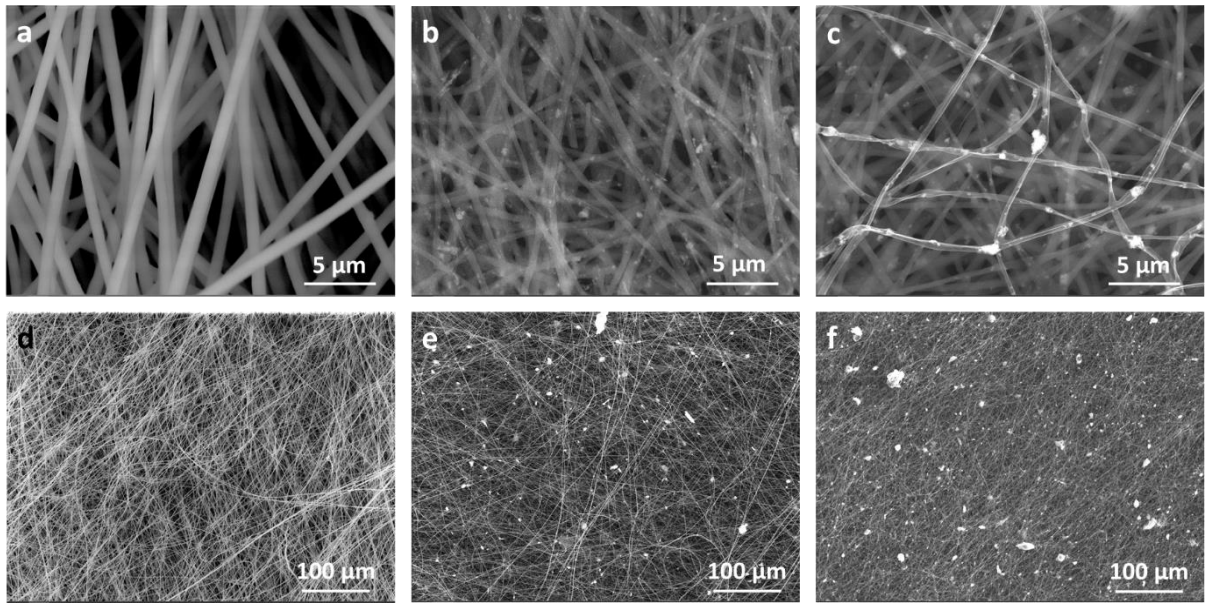

**Figure 2** – Enlarged PAN NFs at 20 k× (a) without nanoparticles, (b) with γ-Fe$_2$O$_3$, (c) with MnZn-Ferrite. PAN NFs at 1000× (d) without nanoparticles, (e) with γ-Fe$_2$O$_3$, (f) with MnZn-Ferrite

### 3.3. Molecular Structure

FTIR spectroscopy in transmittance mode involves passing an infrared light beam through a sample and measuring the intensity of the transmitted light across different wavelengths. Figure 3 compares the three investigated samples; the main peaks from the PAN samples were marked. A noticeable peak appearance was marked at 2243 and 1450 cm$^{-1}$, which are attributed to the highly polar nitrile group –C≡N of PAN by the inclusion of nanopowders and is more distinctive for the MnZn Ferrite due to more interaction [55-58]. The rest of the peaks at 2926 and 1664 cm$^{-1}$ are associated with the –CH$_2$– and C=O groups in the amide structure by undergoing the partial transformation of nitrile groups to amide groups (–CONH$_2$), respectively [55]. As the peak appearing at 2360 cm$^{-1}$ was also detected in our previous work [54], a possible electrostatic interaction between the partial positive charge of metal oxide nanoparticles and the negative charge on the nitrogen of the cyano groups (C≡N) of the PAN, as described in [59]. However, the peaks at ~650 cm$^{-1}$ and 2360 cm$^{-1}$ present only in the NP-containing samples were also attributed to the atmospheric CO$_2$ [60, 61] since the iron oxide present within the composites is a potential adsorbent for CO$_2$ [62].

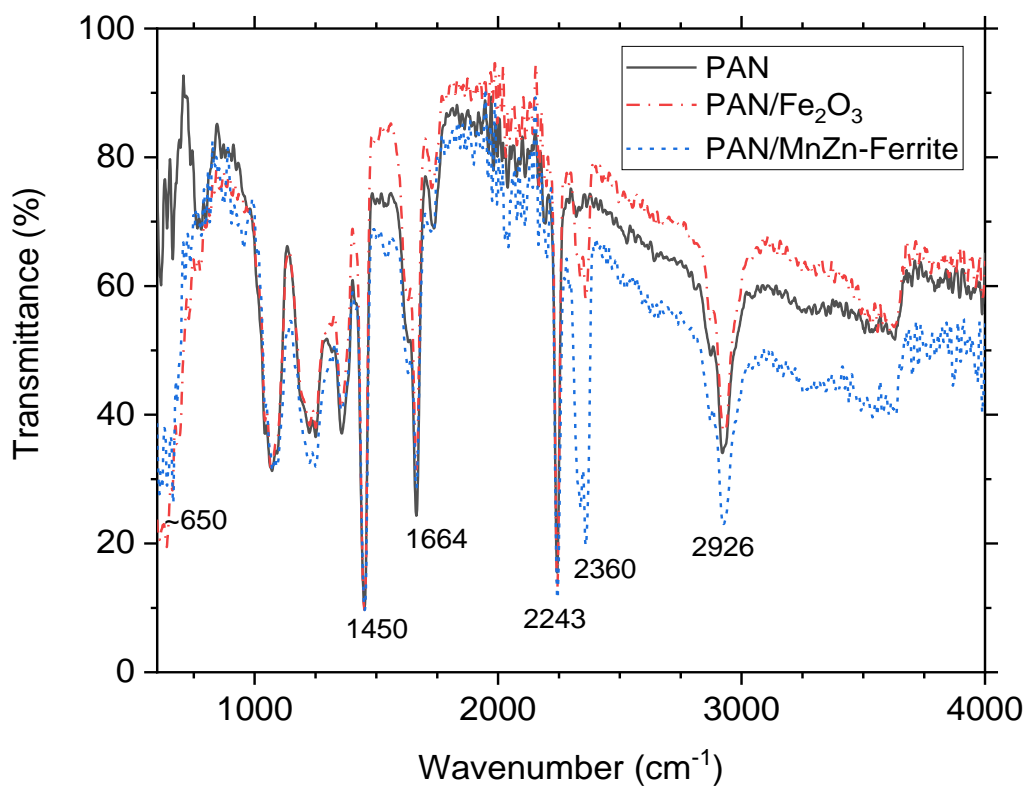

**Figure 3** – Spectroscopic analysis of PAN (solid black), PAN/Fe$_2$O$_3$ (dash-dot red) and PAN/MnZn-Ferrite (dot blue)

### 3.4. Crystalline Structure

Figure 4 shows the structural state of the PAN, PAN/γ-Fe$_2$O$_3$ and PAN/MnZn-Ferrite NF composites. The broad signals at 14.2° and 26.0° confirm the amorphous PAN structure [63]. The peaks at ~35.5°, ~41.6°, and ~50.8° correspond to (220), (311) and (400) planes of the γ-Fe$_2$O$_3$ and MnZn-Ferrite peaks, respectively [64, 65]. In comparison to the largest broad peak position of the PAN, a minor displacement of 0.3° in the peak positions of the NP-containing samples towards higher angles can be attributed to the immediate polymer cooling electrospun on the aluminum foil, resulting in polymer contraction and, hence, a decrease in lattice spacing.

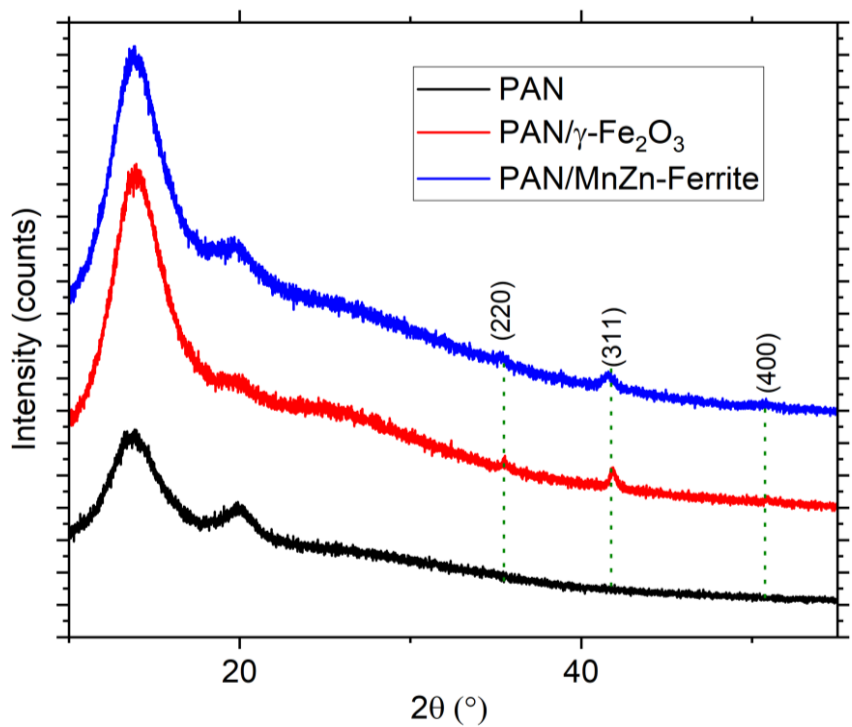

**Figure 4** – Structural differences between the considered samples assessed by XRD

### 3.5. Surface Composition and Chemical State

XPS confirmed the structure of PAN. Figure 5a displays the survey spectrum with strong signals of C1s, N1s and O1s. We also observed some contamination by sulfur, silicon, and phosphorus from the processing of NFs. These elements were not included in Table S1 since we show results from high-resolution spectra, and we have not performed high resolution from these regions corresponding to the mentioned elements. C1s signals at ca 285.6 eV corresponding to –CH$_2$– group in PAN (labeled as C1 in Figure 5b), at ca 286.4 eV corresponding to –CH– in PAN (labeled as C2 in Figure 5b) and –C≡N at ca 287.0 eV confirming the structure of PAN[66]. There is also some adventitious carbon at ca 284.9 eV and a small portion of sp$^2$ at ca 284.2 eV. PAN NFs before heat treatment exhibits only small oxidation, which represents only 2.6 at.% of oxygen and only 2.6 at.% of pyrrolic nitrogen (Figure 5c, Table S1). PAN NFs with NPs before heat treatment showed only slightly higher oxidation, ca. 3-3.2 at.%, and the NPs are fully covered by PAN. We detected only noisy spectra with the only indication of the presence of Fe$_2$O$_3$ (Figure S1a) in the case of the PAN-Fe$_2$O$_3$ sample. In the case of PAN/MnZn-Ferrite there was not even this indication (Figure S1 c,e,f). XPS results of NFs after heat treatment (labeled as "(HT)") indicate an increase in PAN oxidation, and an increase in graphitic carbon (from ca 7 at.% to ca 16 at.%) is observed (Figure 5b, Table S1) that is linked to the interaction of hydrogen bonds. After heat treatment, we detected signals coming from NPs, which can be the result of changes in PAN structure (Table S1, Figure S1 b, d, g, h). This is also accompanied by the changes in nitrogen chemistry, and the N1s signal at ca 398.8 eV indicates the presence of imine or pyridinic structures. N1s at ca 400.6 eV indicate the presence of imide or pyrrolic nitrogen during heat treatment, leading to cyclization-induced C=N bond formation [54]. In the presence of $Fe_2O_3$, it might be correlated to the interaction of nitrile (N1s at ca 399.6 eV) with metal ions ($Fe^{3+}$, $Mn^{4+}$, $Zn^{2+}$). Still, the quantity of this N1s (III) signal is much higher (4.3-4.5 at. %) when compared to the signal of metal ions (0.7 - 1 at.%, Table S1). Most probably, this signal at ca 400.7 eV accounts for the new functional groups of nitrogen on the surface. The higher oxidation is also clear from the increase of oxygen to ca 9.1 – 11.6 at.%. This oxidation after heat treatment is very similar, only slightly higher for PAN (ca 11.6 at.%) than for PAN-containing NPs (9.1-9.6 at.%) (Table S1). The presence of NP was confirmed by the signal of Fe2p at ca 711.3 eV corresponding to $Fe^{3+}$, Mn2p at ca 641.8 eV corresponding to $Mn^{4+}$ and Zn2p at ca 1022.4 eV corresponding to $Zn^{2+}$ (Figure S1).

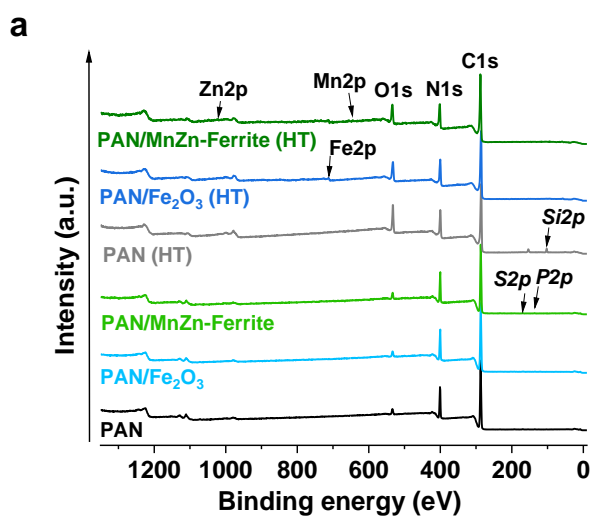

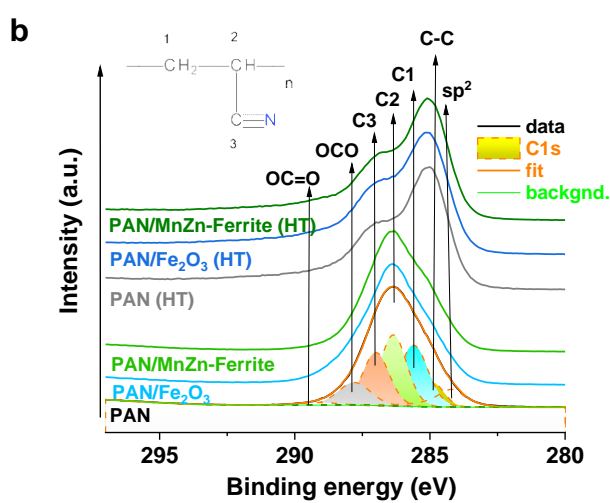

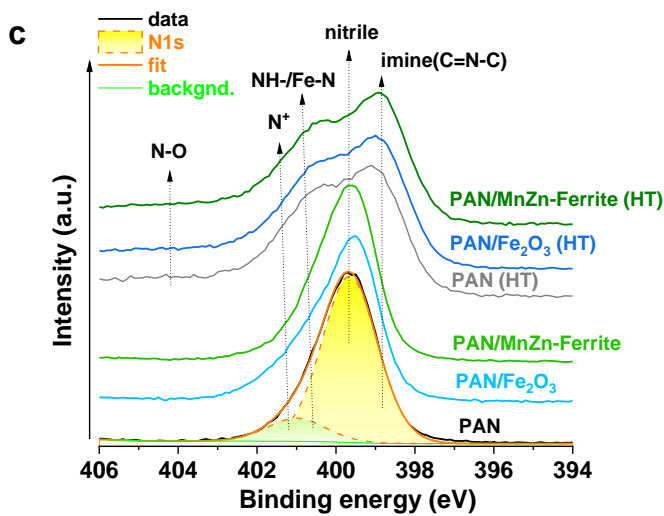

**Figure 5** – X-ray photoelectron spectroscopy (a) survey spectrum, (b) C1s and (c) N1s region of PAN, PAN/Fe$_2$O$_3$ and PAN/MnZn-Ferrite before and after heat treatment

### 3.6. Electrochemical Behaviour

Two different electrolytic mediums indicate that the ionic double-layer behavior differs from each other, as shown by EIS measurements. The studies were implemented in a PBS Buffer (pH = 7.4) and 0.5 M H$_2$SO$_4$. In Figure 6, the absolute impedance values of $|Z|$ vs. PAN/γ-Fe$_2$O$_3$ and PAN/MnZn-Ferrite containing PAN NFs indicate that two electrolytes display the lowest impedance for the PAN/γ-Fe$_2$O$_3$ case at the lowest frequencies. The $|Z|$ values are derived from the Bode Magnitude plots at the 10 mHz frequency, which represents the quasi-DC resistance due to the low frequency, resembling direct current conditions. For the PAN/γ-Fe$_2$O$_3$ in H$_2$SO$_4$ and PAN/MnZn-Ferrite in PBS samples, relatively low impedance values are observed at 10 mHz. These values can be contrasted with the higher resistance observed in the MnZn-Ferrite in both electrolytes, but with an exception after ~1 Hz frequency for PAN/MnZn-Ferrite in H2SO4, indicating the lowest impedance absolute Z value (resistances) due to the nature of the filler and sulfuric acid. The crossover point for PAN/γ-Fe$_2$O$_3$ of the PBS-H$_2$SO$_4$ transition is also ~1 Hz, but this transition is not as sharp as the PAN/MnZn-Ferrite sample.

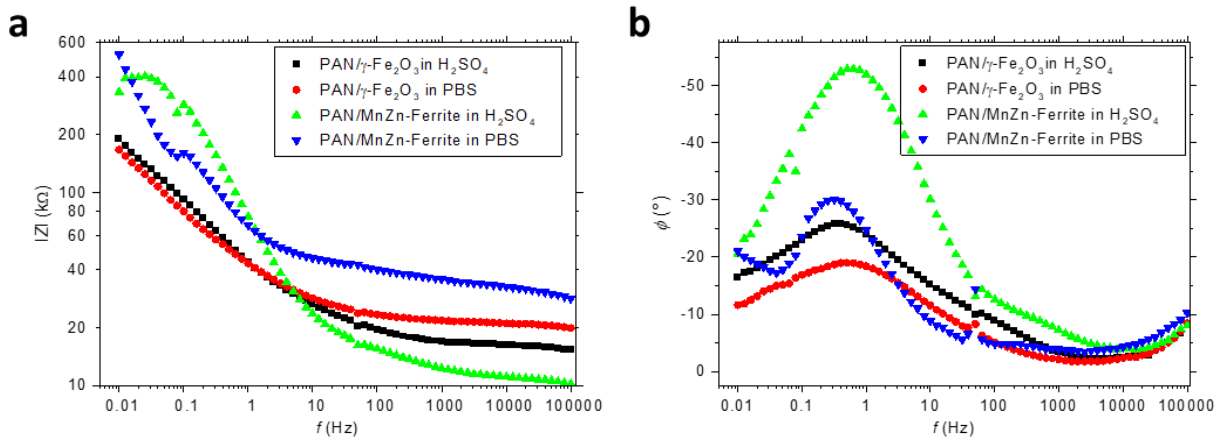

**Figure 6** – (a) Bode magnitude and (b) Bode Phase data in two different electrolytic behaviors within $H_2SO_4$ and PBS

The resistances ($R_s$, $R_1$, $R_2$) in the PBS solution tend to be higher compared to those in the $H_2SO_4$ solution for all samples. PAN/γ-$Fe_2O_3$ sample shows a higher capacitance compared to PAN/MnZn-Ferrite in 0.5 M solutions of $H_2SO_4$ and PBS, referring to the higher amount of ion accumulation in bulk form (Table 1). The PAN/MnZn-Ferrite sample in PBS shows a significantly higher $R_2$ value than in 0.5 M $H_2SO_4$, indicating a higher impedance. The constant phase element ($CPE_1$ and $CPE_2$) values are generally higher in PBS solutions, indicating higher capacitance behavior. The exponents ($n_1$, $n_2$) show variability, with lower values in $H_2SO_4$ solutions, suggesting a more resistive behavior. PAN/MnZn-Ferrite displays lower charge-transfer resistance $R_1$ in both solutions, referring to higher electron transport from the surface to the bulk layers. On the other hand, the surface has more ion exchange characteristics for the PAN/γ-$Fe_2O_3$ sample, as confirmed by the surface resistance $R_2$.

For the bulk-electrolyte interface ($R_1$, $CPE_1$ and $n_1$), a higher capacitance could indicate that more charge is being stored at the interface, which might also mean a higher resistance to charge transfer due to increased interactions at the interface. Nonetheless, the observed inverse relationship between capacitance and resistance in the polarization circuit part ($R_2$, $CPE_2$, and $n_2$) is consistent with electrochemical principles. Higher capacitance suggests improved charge storage and transfer capabilities, which inherently reduce the resistance in the system.

**Table 1** – Equivalent circuit model of the impedances shown in Figure. 6. $R_s$: solution resistance, $CPE_1$ and $CPE_2$: constant phase element exponent associated with bulk layer and polarization, respectively. $R_1$: charge-transfer resistance, $R_2$: polarization resistance, $n_1$ & $n_2$: constant phase element exponent related to bulk layer and polarization, respectively. $\chi^2$: chi-squared error. sample area = 1 cm$^2$.

| 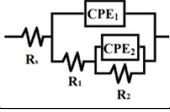 R(Q(R(QR))) | PAN/Fe$_2$O$_3$ in 0.5 M H$_2$SO$_4$ | PAN/MnZn-Ferrite in 0.5 M H$_2$SO$_4$ | PAN/Fe$_2$O$_3$ in 0.5 M PBS | PAN/MnZn-Ferrite in 0.5 M PBS |
|---|---|---|---|---|
| R$_s$/Ω | 1.59*10$^4$ | 1.05*10$^4$ | 2.13*10$^4$ | 3.16*10$^4$ |
| CPE$_1$/Ss$^n$ | 6.31*10$^{-6}$ | 2.40*10$^{-6}$ | 1.01*10$^{-5}$ | 3.41*10$^{-6}$ |
| n$_1$ | 0.53 | 0.57 | 0.55 | 0.44 |
| R$_1$/Ω | 2.38*10$^4$ | 1.07*10$^4$ | 6.99*10$^4$ | 1.74*10$^4$ |
| CPE$_2$/Ss$^n$ | 7.03*10$^{-6}$ | 1.56*10$^{-6}$ | 1.82*10$^{-5}$ | 4.45*10$^{-6}$ |
| n$_2$ | 0.58 | 0.85 | 0.84 | 0.73 |
| R$_2$/Ω | 1.88*10$^5$ | 6.40*10$^5$ | 7.51*10$^4$ | 4.68*10$^5$ |
| Chi-squared / χ2 | 1.02*10$^{-3}$ | 4.65*10$^{-3}$ | 4.31*10$^{-3}$ | 1.20*10$^{-2}$ |

### 3.7. Magnetic Behaviour

Room-temperature magnetic hysteresis loops of the PAN/Fe$_2$O$_3$ (black squares) and PAN/MnZn-Ferrite (red dots) samples are depicted in Figure 7. The M(H) curves have been adjusted based on the mass of the nanocomposite, which includes the electrospun NFs and the magnetic nanoparticles contained therein. Both M(H) curves exhibit a well-defined sigmoidal shape, indicating a smooth magnetization reversal dominated by a rotational process that ends in full magnetic saturation at 10 kOe. The saturation magnetization (M$_s$) values are 6.4 and 2.3 emu/g for the PAN/Fe$_2$O$_3$ and PAN/MnZn-Ferrite samples, respectively. Therefore, the PAN/Fe$_2$O$_3$ sample, which includes only Fe ions in the oxide structure, shows a higher magnetic signal strength compared to PAN/MnZn-Ferrite, which also contains non-magnetic Zn ions. The inset of Figure 7 magnifies the low-field region of the M(H) curves, revealing the hysteretic behavior of both samples. Specifically, the magnetization remanence (M$_r$) values are 1.1 and 0.1 emu/g, while the coercive field (H$_c$) values are 106 and 23 Oe for the PAN/Fe$_2$O$_3$ and PAN/MnZn-Ferrite samples, respectively. Thus, the MnZn-Ferrite nanoparticles exhibit a reduced effective magnetic anisotropy.

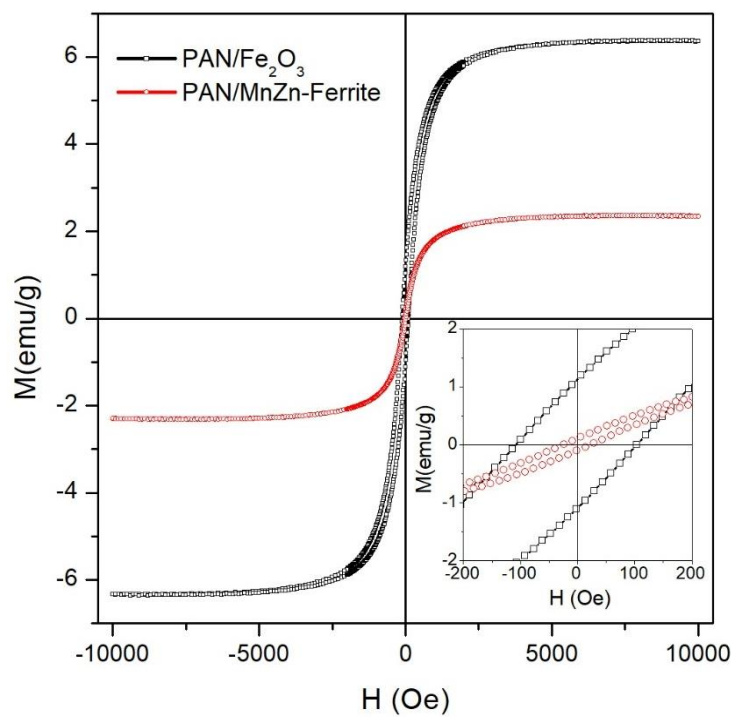

**Figure 7** – Room-temperature magnetic hysteresis loops of the PAN/Fe$_2$O$_3$ (black squares) and PAN/MnZn-Ferrite (red dots) samples; inset: low magnetic field enlargement.

### 3.8. Cytocompatibility

Since Fe$_2$O$_3$/PAN NF and MnZn-Ferrite NPs/PAN NF are intended for biomedical applications, as discussed earlier, it is crucial to ensure their cytocompatibility. The nanoparticles incorporated into PAN may influence cellular responses either positively or negatively, depending on their concentration and interaction with cells. To assess this, cytocompatibility analyses were conducted using the direct method with MSCs.

Figure 8 illustrates the cytocompatibility of MSCs cultivated directly on the samples. A comparison of the cellular metabolic functions of cultivated cells on the materials revealed a significant difference between PAN-MnZnFerrite and both PAN and PAN-Fe$_2$O$_3$ during the first 24 hours ($p < 0.05$).

This significant reduction in cells' metabolic activity on PAN-MnZnFerrite may be attributed to the release of manganese and/or zinc ions and their subsequent uptake by cells. While the exact contribution to cell cytotoxicity remains unclear, some research suggests that these ions could induce mitochondrial dysfunction, potentially impairing cellular metabolism [67-69].

However, at the second time point (48 hours), the significant difference observed earlier disappeared, thus indicating the material's cytocompatibility. These results were further confirmed by the LDH assay, which did not show any significant difference and, consequently, any difference in terms of LDH release between cells cultivated on the control material and the doped samples. The viability of cells is lastly ensured by the Live/Dead assay conducted at the final time point, which showed that all observed cells were alive and displayed a morphology comparable to that of cells grown on the other materials. The limited increase in viability observed between 24 and 48 hours on pure PAN nanofibers reflects the material's limited bioactivity and lack of biochemical cues for hMSC proliferation. This behavior is in contrast to the MnZn ferrite and $Fe_2O_3$-loaded PAN samples, which showed enhanced biological response supported by multiple assays.

In summary, the Alamar Blue assay showed only modest changes for PAN between 24 h and 48 h, the LDH assay at 48 h confirmed the absence of cytotoxicity. Furthermore, the Live/Dead imaging showed viable cells with normal morphology.

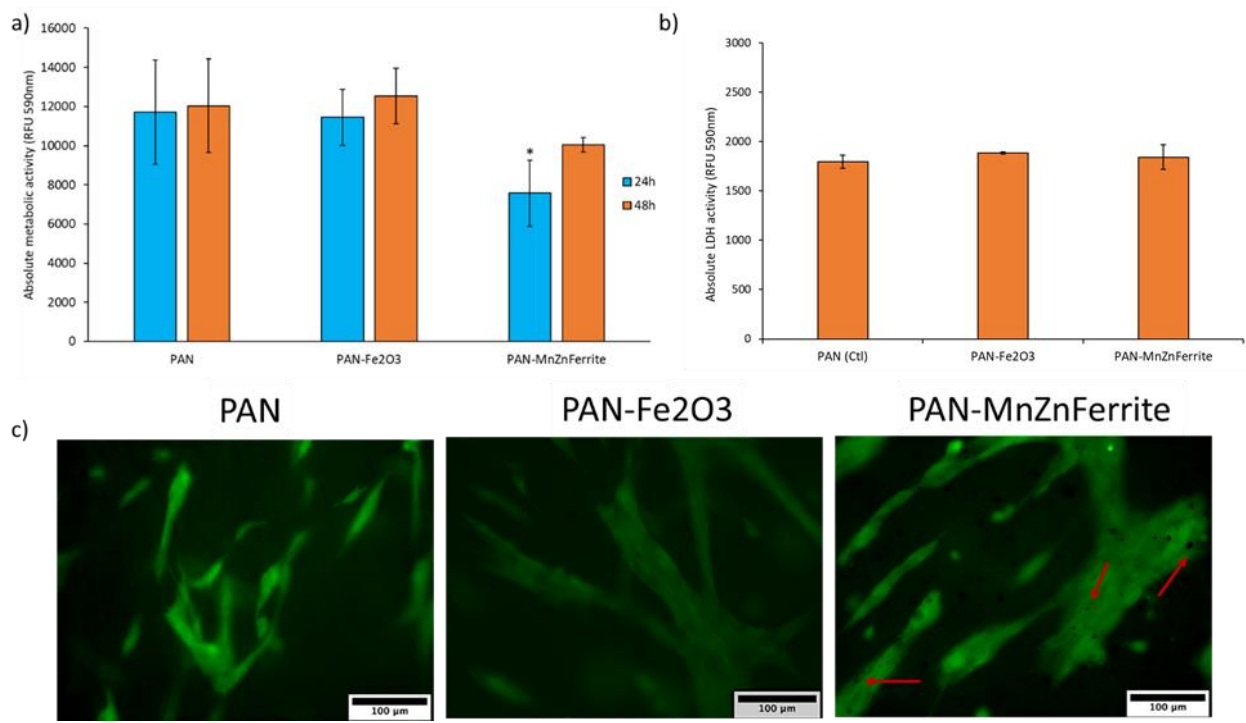

**Figure 8** – Direct cytocompatibility evaluation of PAN NF, PAN/Fe2O3, and PAN/MnZn-Ferrite samples. hMSCs were cultured directly on the sample surfaces for 24 and 48 hours, and cytocompatibility was assessed through the following: (a) absolute metabolic activity of hBMSCs, (b) Absolute LDH activity measured after 48 hours in hMSCs cultured on unloaded (PAN, control) and loaded materials (PAN–$Fe_2O_3$ and PAN–MnZn ferrite); (c) Live/Dead imaging after 48 hours of cultivation. *$p < 0.05$ vs. PAN and PAN/$Fe_2O_3$.

SEM images in Figure 9a confirm proper cell adhesion, morphology, and spreading on the nanofibers. Notably, a higher cell density is observed on both doped materials compared to the control. The zoomed-in images further highlight distinct points of contact between the cells and the doped materials, demonstrating the scaffolds' ability to create a favorable environment for cell attachment and growth. Figures 9b and 9c show that the expression levels of Cadherin-1 and Ki67—key markers of cell adhesion and proliferation, respectively—are significantly higher in cells cultured on PAN-MnZnFerrite and PAN-$Fe_2O_3$ than in those cultured on undoped PAN (**$p < 0.01$ and *$p < 0.05$, respectively). Previous studies suggest that both iron oxide nanoparticles and MnZn ferrite composites can enhance cell adhesion and proliferation on electrospun substrates by increasing the surface roughness of the NFs, promoting protein adsorption, and improving the hydrophilicity of the resulting scaffold [70, 71]. Figure 9d reports the results related to the Click it EdU proliferation assay. Results demonstrated that cell proliferation is preserved in all the materials and that there are no significant differences among the groups. The discrepancy between Click-iT and Ki67 results (panel c) may reflect differences in cell cycle distribution. Indeed, the Click-iT detects only cells in the S phase of the cell cycle, while Ki67 indicates overall proliferative activity, including G1, G2, and M phases [72, 73].

Inclusion of $MnZnFe_2O_4$ nanoparticles in PAN NFs enhances hMSC adhesion and spreading compared with $Fe_2O_3$ at the same loading. The different ionic radii of $Mn^{2+}$, $Zn^{2+}$, and $Fe^{3+}$ mean the surface can present a more diverse array of potential interaction sites and variable surface charge for protein ligands [74, 75]. Furthermore, the controlled release of $Mn^{2+}$ and $Zn^{2+}$ ions—both known to support osteogenic differentiation—further stimulates hMSC proliferation and lineage commitment, effects not achievable with the simpler $Fe^{3+}$-dominated surface of $Fe_2O_3$ [76, 77].

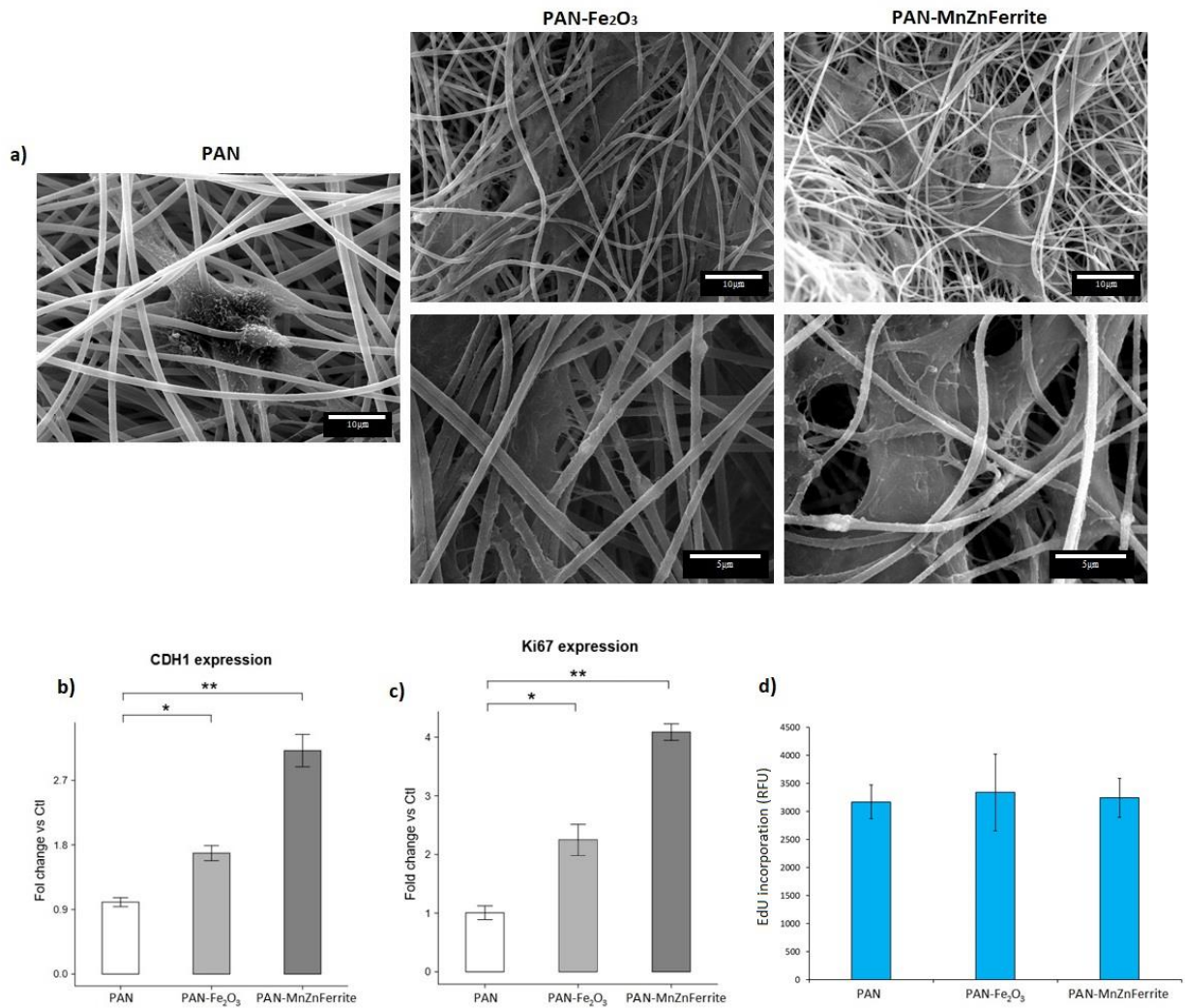

**Figure 9** – Cell adhesion and proliferation. a) Representative images showing hMSC attachment on the sample surfaces after 48 hours of incubation. b-c) Relative expression of Cadherin-1 (CDH1) and Ki67 in hMSCs cultured on PAN/Fe₂O₃ and PAN/MnZn Ferrite compared to PAN (control); d) EdU-based proliferation assay. *$p<0.05$ **$p<0.01$

4. Conclusions

Advanced characterization via spectroscopic (XPS and FTIR), crystallographic (XRD), magnetic (AGFM), morphologic (SEM) and cytotoxicity analysis helped to better understand the presence of oxide particles and the mechanism of interaction between magnetic metal oxide nanoparticles and polymer matrix. Homogeneously dispersed metal oxide in polymer solution showcases notable physical traits like uniformity in electrospun NF diameter and metal oxide particle dispersion on these NFs. Advanced characterization techniques, XPS confirmed the structure of PAN NFs and revealed increased oxidation and graphitic carbon content after heat treatment. New nitrogen functionalities and interactions between the polar nitrile groups of PAN and $Fe_2O_3$ and MnZn Ferrite nanoparticles was confirmed. MnZn Ferrite/PAN NFs demonstrates smaller magnetic remanence and coercivity compared to $Fe_2O_3$/PAN referring to reduced magnetic anisotropy. Absolute metabolic and LDH activity tests demonstrated that Mn-Zn ferrite and iron oxide nanoparticles loaded onto PAN NFs remarkably enhanced the viability or conformation of human mesenchymal stromal cells in comparison to the reference PAN sample, making them a promising materials for biomedical applications. As compared to PAN and PAN/$Fe_2O_3$ NFs, Cadherin-1 and Ki67—core identifiers of cell adhesion and proliferation, respectively—are remarkably higher in cells cultured on PAN/MnZn Ferrite, as also evidenced by SEM images revealing hMSC attachment on the sample surfaces after 48 hours of incubation. These positive, albeit preliminary, biological findings, combined with the intrinsic magnetic properties of the incorporated nanoparticles, suggest the potential for this material to be exploited in the field of magnetic resonance imaging, targeted drug administration, and magnetic hyperthermia for oncology, as well as other advanced therapeutic and diagnostic applications.

**Acknowledgments** This work is dedicated to the memory of Prof. Dr. Maria Omastova, whose invaluable contributions were essential to the success of this collaboration. A distinguished scholar in polymers, polymeric composites, and XPS characterization, Prof. Dr. Omastova passed away on May 9, 2024, before this paper's publication. Her expertise and dedication to international scientific collaboration will be deeply missed.

**Funding** This research was partially funded by the Slovak agency VEGA grant number 02/006/22. The authors would like to acknowledge the contribution of the COST Action CA21155 (HISTRATE).

**Supplementary Information** The online version contains supplementary material at

**Competing Interest** The authors declare that they have no known competing financial interests or personal relationships that could have appeared to influence the work reported in this paper.

**Data Availability** All data is presented in the figures, graphics and tables in the context of the manuscript.

**References**

[1] A. Heymer, D. Haddad, M. Weber, U. Gbureck, P.M. Jakob, J. Eulert, U. Nöth, Biomaterials, 29, 1473-1483 (2008), http://doi.org/10.1016/j.biomaterials.2007.12.003.

[2] N. Wang, Y. Xie, Z. Xi, Z. Mi, R. Deng, X. Liu, R. Kang, X. Liu, Front Bioeng Biotechnol, 10, (2022), http://doi.org/10.3389/fbioe.2022.937803.

[3] L.M. Gradinaru, S. Vlad, R.C. Ciobanu, Membranes, 12, 1127 (2022), http://doi.org/10.3390/membranes12111127.

[4] H. Sadeghzadeh, H. Dianat-Moghadam, A.R. Del Bakhshayesh, D. Mohammadnejad, A. Mehdipour, Stem Cell Research & Therapy, 14, 194 (2023), http://doi.org/10.1186/s13287-023-03426-0.

[5] K. Kizilbey, E.N. Koprulu, H. Temur, S. Canim Ates, S. Ozer, Materials, 17, 6064 (2024), http://doi.org/10.3390/ma17246064.

[6] P. Hu, J. Lu, C. Li, Z. He, X. Wang, Y. Pan, L. Zhao, ACS Appl Bio Mater, 7, 1569-1578 (2024), http://doi.org/10.1021/acsabm.3c01074.

[7] J. Nowak-Jary, A. Plociennik, B. Machnicka, Int J Mol Sci, 25, 9098 (2024), http://doi.org/10.3390/ijms25169098.

[8] M. Kasprzyk, G. Opila, A. Hinz, S. Stankiewicz, M. Bzowska, K. Wolski, J. Dulinska-Litewka, J. Przewoznik, C. Kapusta, A. Karewicz, ACS Appl Mater Interfaces, 17, 9059-9073 (2025), http://doi.org/10.1021/acsami.4c20101.

[9] L. Maldonado-Camargo, M. Unni, C. Rinaldi, Methods Mol Biol, 1570, 47-71 (2017), http://doi.org/10.1007/978-1-4939-6840-4_4.

[10] Q. Wang, B. Chen, F. Ma, S. Lin, M. Cao, Y. Li, N. Gu, Nano Res, 10, 626-642 (2017), http://doi.org/10.1007/s12274-016-1322-4.

[11] Nowak-Jary J, M. B, Int J Nanomed, 18, 4067-4100 (2023), http://doi.org/10.2147/IJN.S415063.

[12] S. Kargozar, F. Baino, S. Hamzehlou, M.R. Hamblin, M. Mozafari, Chemical Society Reviews, 49, 5008-5057 (2020), http://doi.org/10.1039/c8cs01021h.

[13] A. Shradhanjali, J.T. Wolfe, B.J. Tefft, Tissue Eng Part B Rev, (2024), http://doi.org/10.1089/ten.TEB.2024.0103.

[14] H.O. Alsaab, A.S. Al-Hibs, R. Alzhrani, K.K. Alrabighi, A. Alqathama, A. Alwithenani, A.H. Almalki, Y.S. Althobaiti, International Journal of Molecular Sciences, 22, 1631 (2021), http://doi.org/10.3390/ijms22041631.

[15] O. Agboola, O.S.I. Fayomi, A. Ayodeji, A.O. Ayeni, E.E. Alagbe, S.E. Sanni, E.E. Okoro, L. Moropeng, R. Sadiku, K.W. Kupolati, B.A. Oni, Membranes, 11, 139 (2021), http://doi.org/10.3390/membranes11020139.

[16] A. Luzio, E.V. Canesi, C. Bertarelli, M. Caironi, Materials, 7, 906-947 (2014), http://doi.org/10.3390/ma7020906.

[17] S. Ma, A. Li, L. Pan, Polymers, 16, 2459 (2024), http://doi.org/10.3390/polym16172459.

[18] S.M.S. Shahriar, J. Mondal, M.N. Hasan, V. Revuri, D.Y. Lee, Y.-K. Lee, Nanomaterials, 9, 532 (2019), http://doi.org/10.3390/nano9040532.

[19] T. Maeda, Y.-J. Kim, T. Aoyagi, M. Ebara, Fibers, 5, 13 (2017), http://doi.org/10.3390/fib5010013.

[20] D.T. Govindaraju, H.H. Kao, Y.M. Chien, J.P. Chen, Int J Mol Sci, 25, 9803 (2024), http://doi.org/10.3390/ijms25189803.

[21] S. Pisani, A. Piazza, R. Dorati, I. Genta, M. Rosalia, E. Chiesa, G. Bruni, R. Migliavacca, B. Conti, Int J Pharm, 125393 (2025), http://doi.org/10.1016/j.ijpharm.2025.125393.

[22] Z. Moazzami Goudarzi, A. Zaszczyńska, T. Kowalczyk, P. Sajkiewicz, Pharmaceutics, 16, 93 (2024), http://doi.org/10.3390/pharmaceutics16010093.

[23] D. Coetzee, M. Venkataraman, J. Militky, M. Petru, Polymers, 12, 742 (2020), http://doi.org/10.3390/polym12040742.

[24] Y.R.V. Shih, C.N. Chen, S.W. Tsai, Y.J. Wang, O.K. Lee, Stem Cells, 24, 2391-2397 (2006), http://doi.org/10.1634/stemcells.2006-0253.

[25] R.A. Neal, S.G. McClugage, M.C. Link, L.S. Sefcik, R.C. Ogle, E.A. Botchwey, Tissue Eng, Part C, 15, 11-21 (2009), http://doi.org/10.1089/ten.tec.2007.0366.

[26] J.M. Coburn, M. Gibson, S. Monagle, Z. Patterson, J.H. Elisseeff, Proc Natl Acad Sci U S A, 109, 10012-10017 (2012), http://doi.org/doi:10.1073/pnas.1121605109.

[27] Y.M. Kolambkar, A. Peister, A.K. Ekaputra, D.W. Hutmacher, R.E. Guldberg, Tissue Eng, Part A, 16, 3219-30 (2010), http://doi.org/10.1089/ten.TEA.2010.0004.

[28] H.-H. Kao, D.T. Govindaraju, B.S. Dash, J.-P. Chen, J Compos Sci, 9, 31 (2025), http://doi.org/10.3390/jcs9010031.

[29] D. Jhala, H.A. Rather, R. Vasita, Biomed Mater, 15, 035011 (2020), http://doi.org/10.1088/1748-605X/ab772e.

[30] D. Yu, J. Wang, K.-j. Qian, J. Yu, H.-y. Zhu, J Zhejiang Univ Sci B, 21, 871-884 (2020), http://doi.org/10.1631/jzus.B2000355.

[31] T.A. Adegbola, O. Agboola, O.S.I. Fayomi, Results in Engineering, 7, 100144 (2020), http://doi.org/10.1016/j.rineng.2020.100144.

[32] X. Chen, Y. Su, F. Shen, Y. Wan, Journal of Membrane Science, 384, 44-51 (2011), http://doi.org/10.1016/j.memsci.2011.09.002.

[33] T. Blachowicz, A. Ehrmann, Materials, 13, 152 (2020), http://doi.org/10.3390/ma13010152.

[34] M.Q. Khan, M.A. Alvi, H.H. Nawaz, M. Umar, Nanomaterials, 14, 1305 (2024), http://doi.org/10.3390/nano14151305.

[35] X. Li, L. Wang, S. Li, S. Yu, Z. Liu, Q. Liu, X. Dong, Int J Biol Macromol, 274, 133381 (2024), http://doi.org/10.1016/j.ijbiomac.2024.133381.

[36] Q.-Y. Wu, X.-N. Chen, L.-S. Wan, Z.-K. Xu, The Journal of Physical Chemistry B, 116, 8321-8330 (2012), http://doi.org/10.1021/jp304167f.

[37] G. Henrici-Olivé, S. Olivé. Molecular interactions and macroscopic properties of polyacrylonitrile and model substances. In: Chemistry. Berlin, Heidelberg: Springer Berlin Heidelberg; 1979. p. 123–152.

[38] J. Yin, V.S. Reddy, A. Chinnappan, S. Ramakrishna, L. Xu, Polymer Reviews, 63, 715-762 (2023), http://doi.org/10.1080/15583724.2022.2158467.

[39] Y. Yang, B. Xu, Y. Gao, M. Li, Acs Applied Materials & Interfaces, 13, 49927-49935 (2021), http://doi.org/10.1021/acsami.1c14273.

[40] R. Mincheva, O. Stoilova, H. Penchev, T. Ruskov, I. Spirov, N. Manolova, I. Rashkov, European Polymer Journal, 44, 615-627 (2008), http://doi.org/10.1016/j.eurpolymj.2007.11.001.

[41] D. Pathak, A. Sharma, D.P. Sharma, V. Kumar, Applied Surface Science Advances, 18, 100471 (2023), http://doi.org/10.1016/j.apsadv.2023.100471.

[42] R. Zhao, X. Li, Y. Li, Y. Li, B. Sun, N. Zhang, S. Chao, C. Wang, Journal of Colloid and Interface Science, 505, 1018-1030 (2017), http://doi.org/10.1016/j.jcis.2017.06.094.

[43] W. Xu, W. Zheng, F. Wang, Q. Xiong, X.-L. Shi, Y.K. Kalkhajeh, G. Xu, H. Gao, Chemical Engineering Journal, 403, 126349 (2021), http://doi.org/10.1016/j.cej.2020.126349.

[44] G.S. Parkinson, Surface Science Reports, 71, 272-365 (2016), http://doi.org/10.1016/j.surfrep.2016.02.001.

[45] A. Mahapatra, N. Garg, B.P. Nayak, B.G. Mishra, G. Hota, Journal of Applied Polymer Science, 124, 1178-1185 (2012), http://doi.org/10.1002/app.35076.

[46] C. Barozzi, G. Cairo, R. Fumero, S. Scuri, M.C. Tanzi, P. Albonico, Life Support Syst, 3 Suppl 1, 490-4 (1985), http://doi.org/10.1016/0376-7388(88)80019-X.

[47] A. Mamun, L. Sabantina, Polymers, 15, 1902 (2023), http://doi.org/10.3390/polym15081902.

[48] M.R. Ramezani, Z. Ansari-Asl, E. Hoveizi, A.R. Kiasat, Materials Chemistry and Physics, 229, 242-250 (2019), http://doi.org/10.1016/j.matchemphys.2019.03.031.

[49] L.M. Bauer, S.F. Situ, M.A. Griswold, A.C.S. Samia, Nanoscale, 8, 12162-12169 (2016), http://doi.org/10.1039/C6NR01877G.

[50] C. Pucci, A. Degl'Innocenti, M. Belenli Gümüş, G. Ciofani, Biomaterials Science, 10, 2103-2121 (2022), http://doi.org/10.1039/D1BM01963E.

[51] E.M. Materón, C.M. Miyazaki, O. Carr, N. Joshi, P.H.S. Picciani, C.J. Dalmaschio, F. Davis, F.M. Shimizu, Applied Surface Science Advances, 6, 100163 (2021), http://doi.org/10.1016/j.apsadv.2021.100163.

[52] K. Huner, B. Sarac, E. Yüce, A. Rezvan, M. Micusik, M. Omastova, J. Eckert, A.S. Sarac, Mol Syst Des Eng, 8, 394-406 (2023), http://doi.org/10.1039/D2ME00181K.

[53] S. Gumrukcu, V. Soprunyuk, B. Sarac, E. Yüce, J. Eckert, A.S. Sarac, Mol Syst Des Eng, 6, 476–492 (2021), http://doi.org/10.1039/d1me00008j.

[54] B. Sarac, V. Soprunyuk, G. Herwig, S. Gümrükçü, E. Kaplan, E. Yüce, W. Schranz, J. Eckert, L.F. Boesel, A.S. Sarac, Nanoscale Adv, 6, 6184-6195 (2024), http://doi.org/10.1039/d4na00631c.

[55] Q. Liu, N. Xu, L. Fan, A. Ding, Q. Dong, Chem Eng Sci, 228, 115993 (2020), http://doi.org/10.1016/j.ces.2020.115993.

[56] W. Li, Z. Yang, G. Zhang, Q. Meng, Ind Eng Chem Res, 52, 6492-6501 (2013), http://doi.org/10.1021/ie303122u.

[57] J. Gao, X. Wang, J. Zhang, R. Guo, Sep Purif Technol, 159, 116-123 (2016), http://doi.org/10.1016/j.seppur.2016.01.005.

[58] G. Zhang, H. Yan, S. Ji, Z. Liu, J Membr Sci, 292, 1-8 (2007), http://doi.org/10.1016/j.memsci.2006.11.023.

[59] S. Nag, S. Pradhan, D. Das, B. Tudu, R. Bandyopadhyay, R.B. Roy, IEEE Sens J, 22, 42-49 (2022), http://doi.org/10.1109/JSEN.2021.3128520.

[60] E. Pullicino, W. Zou, M. Gresil, C. Soutis, Appl Compos Mater, 24, 301-311 (2017), http://doi.org/10.1007/s10443-016-9559-3.

[61] M. Aliahmad, N. Nasiri Moghaddam, Mater Sci-Pol, 31, 264-268 (2013), http://doi.org/10.2478/s13536-012-0100-6.

[62] S. Agarwal, M.P. Mudoi, S. Singhal, A.S. Khichi, A. Dhyani. 41 - Ferrites and Fe oxides as effective materials for the removal of $CO_2$. In: Pal Singh J, Chae KH, Srivastava RC, Caltun OF, editors. Ferrite Nanostructured Magnetic Materials: Woodhead Publishing; 2023, p. 831-851.

[63] Y. Zhao, Z. Zhao, M. Wei, X. Jiang, H. Li, J. Gao, L. Hou, Prog Nat Sci: Mater Int, 28, 337-344 (2018), http://doi.org/10.1016/j.pnsc.2018.04.013.

[64] H.A. Alshamsi, B.S. Hussein, Asian J Chem, 30, 273-279 (2017), http://doi.org/10.14233/ajchem.2018.20888.

[65] D. Arcos, R. Valenzuela, M. Vallet-Regí, M. Vázquez, J Mater Res, 14, 861-865 (1999), http://doi.org/10.1557/JMR.1999.0115.

[66] G. Beamson, D.R. Briggs. The XPS of polymers database. In: Surface Spectra Ltd.; 2000.

[67] S. Brown, N.L. Taylor, Environ Toxicol Pharmacol, 7, 49-57 (1999), http://doi.org/10.1016/S1382-6689(98)00054-4.

[68] J. Diessl, J. Berndtsson, F. Broeskamp, L. Habernig, V. Kohler, C. Vazquez-Calvo, A. Nandy, C. Peselj, S. Drobysheva, L. Pelosi, F.N. Vögtle, F. Pierrel, M. Ott, S. Büttner, Nat Commun, 13, 6061 (2022), http://doi.org/10.1038/s41467-022-33641-x.

[69] O.R.M. Bagshaw, R. Alva, J. Goldman, J.W. Drelich, J.A. Stuart. 33 - Mitochondrial zinc toxicity. In: de Oliveira MR, editor. Mitochondrial Intoxication: Academic Press; 2023, p. 723-744.

[70] C.J. Mortimer, C.J. Wright, Biotechnol J, 12, (2017), http://doi.org/10.1002/biot.201600693.

[71]     Y. Wang, Y. Miao, G. Li, M. Su, X. Chen, H. Zhang, Y. Zhang, W. Jiao, Y. He, J. Yi, X. Liu, H. Fan, Mater Today Adv, 8, 100119 (2020), http://doi.org/10.1016/j.mtadv.2020.100119.

[72]     R.C. Leif, J.H. Stein, R.M. Zucker, Cytometry Part A, 58A, 45-52 (2004), http://doi.org/10.1002/cyto.a.20012.

[73]     I. Miller, M. Min, C. Yang, C. Tian, S. Gookin, D. Carter, S.L. Spencer, Cell Rep, 24, 1105-1112.e5 (2018), http://doi.org/10.1016/j.celrep.2018.06.110.

[74]     Y.J. Kim, B.C. Park, Y.S. Choi, M.J. Ko, Y.K. Kim, Electron Mater Lett, 15, 471-480 (2019), http://doi.org/10.1007/s13391-019-00141-y.

[75]     R. Augustine, A. Hasan, R. Primavera, R.J. Wilson, A.S. Thakor, B.D. Kevadiya, Mater Today Commun, 25, 101692 (2020), http://doi.org/10.1016/j.mtcomm.2020.101692.

[76]     Y.J. Kim, J. Lee, G.B. Im, J. Song, J. Song, J. Chung, T. Yu, S.H. Bhang, Materials, 14, 412 (2021), http://doi.org/10.3390/ma14020412.

[77]     Z. Du, H. Leng, L. Guo, Y. Huang, T. Zheng, Z. Zhao, X. Liu, X. Zhang, Q. Cai, X. Yang, Composites, Part B, 190, 107937 (2020), http://doi.org/10.1016/j.compositesb.2020.107937.

Supplementary Material

# Enhanced Mesenchymal Stem Cell Response with Preserved Biocompatibility via (MnZn)Ferrite–Polyacrylonitrile Composite Nanofiber Membranes

Baran Sarac[a,b]*, Elham Sharifikolouei[c], Matej Micusik[d], Alessandro Scalia[c], Ziba Najmi[c], Andrea Cochis[c], Lia Rimondini[c], Gabriele Barrera[e], Marco Coisson[e], Selin Gümrükcü[f], Eray Yüce[a,g] and A. Sezai Sarac[h]

[a] Erich Schmid Institute of Materials Science, Austrian Academy of Sciences (ÖAW), 8700 Leoben, Austria
[b] Chair of Casting Research, Montanuniversität Leoben, 8700 Leoben, Austria
[c] Università del Piemonte Orientale UPO, Department of Health Sciences, Center for Translational Research on Autoimmune and Allergic Diseases-CAAD, 28100 Novara, Italy

[d] Polymer Institute, Slovak Academy of Sciences, Dubravska cesta 9, Bratislava 84541, Slovakia

[e] Istituto Nazionale di Ricerca Metrologica (INRiM), Strada delle Cacce 91, Torino, 10135 Italy

[f] Department of Chemistry, Istanbul Technical University, 34469 Istanbul, Türkiye

[g] Department of Materials Science, Chair of Materials Physics, Montanuniversität Leoben, 8700 Leoben, Austria

[h] Polymer Science & Technology, Istanbul Technical University, Maslak, 34469 Istanbul, Turkey

* E-mail: baransarac@gmail.com

**Table S1** – Apparent surface chemical composition of PAN, PAN/Fe$_2$O$_3$ and PAN/MnZnFerrite as determined by XPS. Heat treated samples are denoted by (HT).

PAN

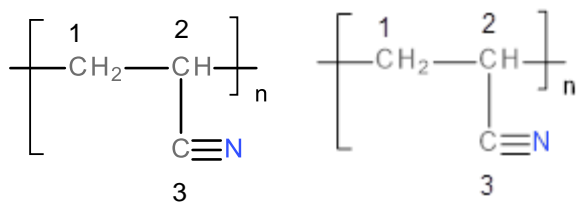

| sample | Surface chemical composition (at.%) | | | |
|---|---|---|---|---|
| | C1s sp$^2$/sp$^3$/C1/ C2/C3/OCO/OC=O | O1s I/II/III | N1s I/II/III/IV/V | Fe2p/Mn2p/ Zn2p |
| PAN | 77.2 7.2/5.8/19.1/20.6/15.4/ 8.1/1.0 | 2.6 0.9/0.9/0.8 | 20.2 -/17.6/2.6/-/- | -/-/- |
| PAN/Fe$_2$O$_3$ | 77.2 7.2/4.4/21.5/15.3/14.4/ 11.9/2.5 | 3.0 1.1/0.8/1.1 | 19.8 -/14.4/5.4/-/- | -/-/- |
| PAN/MnZn-Ferrite | 77.7 7.4/5.5/22.1/16.2/17.4/ 7.4/1.7 | 3.2 1.3/1.0/0.9 | 19.1 -/16.2/2.9/-/- | -/-/- |
| PAN (HT) | 70.9 15.5/13.4/16.7/4.3/7.9/ 10.5/2.6 | 11.6 4.0/5.4/2.2 | 17.5 7.7/3.6/4.6/1.1/0.5 | -/-/- |
| PAN/Fe$_2$O$_3$ (HT) | 71.8 16.2/12.6/16.9/4.2/8.2/ 10.5/3.2 | 9.6 4.8/2.1/2.7 | 17.9 7.5/3.6/4.5/1.5/0.8 | 0.7/-/- |
| PAN/MnZn-Ferrite (HT) | 72.6 15.8/15.4/16.1/4.9/9.0/ 8.4/3.0 | 9.1 4.2/2.3/2.6 | 17.3 7.7/3.0/4.3/1.5/0.8 | 0.6/0.3/0.1 |

O1s: I: C=O/oxides ~ 531 eV, II: C-O ~ 532 eV, III: C-O-C ~ 534 eV
N1s: I: imine (-N=C-C-) ~ 398.8 eV, II: nitrile/pyrrolic (C≡N/-NH-) ~ 399.6 eV, III: Fe-N (charge transfer/imide ~ 400.7 eV, IV: alkyl amonium (NR$_4^+$) ~ 402.0 eV, V: oxidized nitrogen (N-O) ~ 404.5 eV

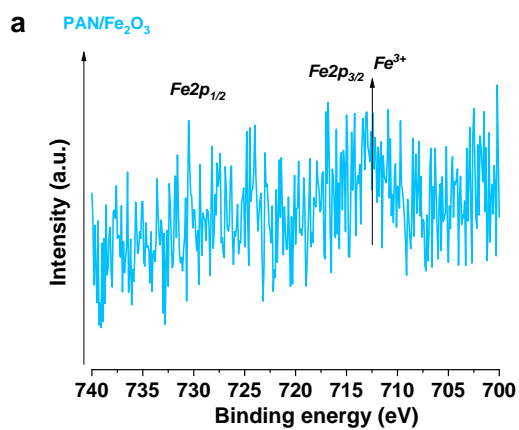
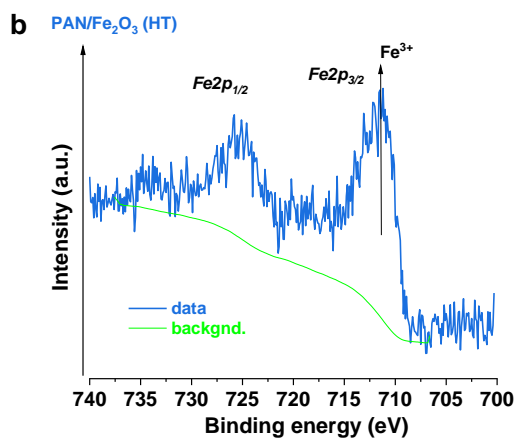
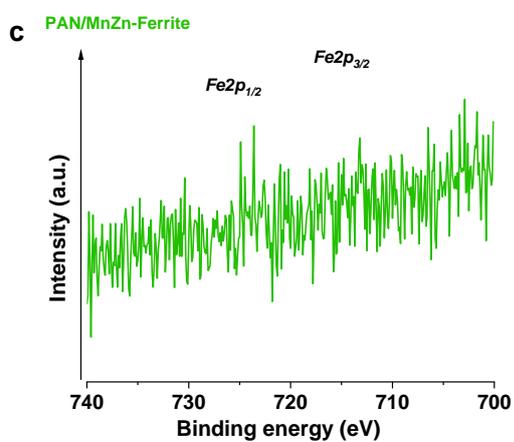
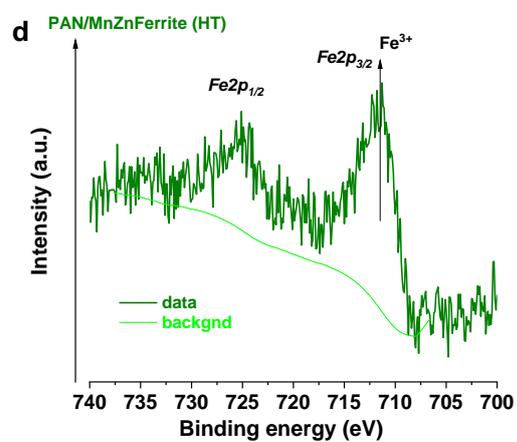
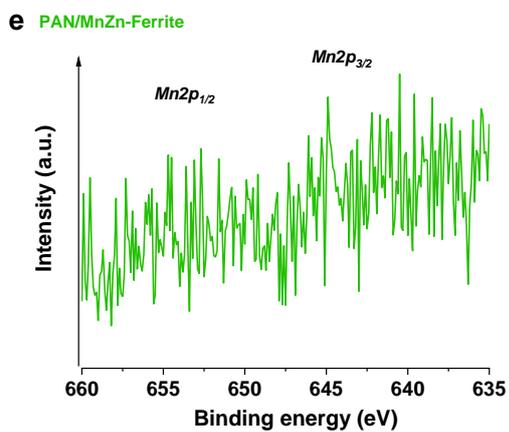
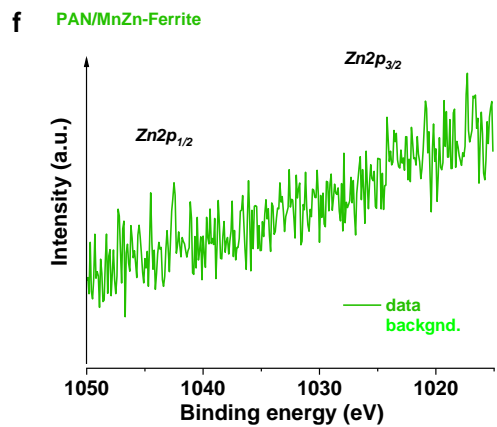

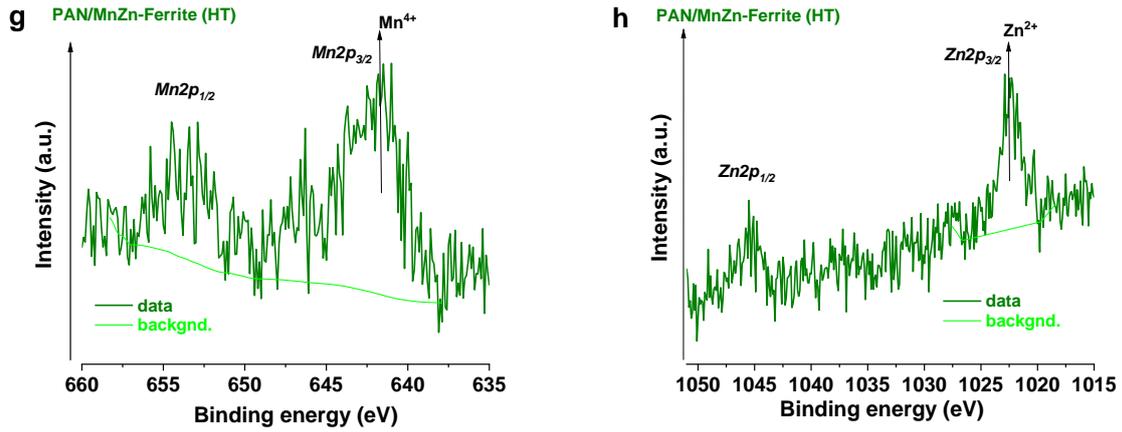

**Figure S1** – XPS of Fe2p region of (a) PAN/Fe$_2$O$_3$, (b) PAN/Fe$_2$O$_3$ (HT), (c) PAN/MnZn-Ferrite, (d) PAN/MnZn-Ferrite (HT); Mn2p region of (e) PAN/MnZn-Ferrite, (f) PAN/MnZn-Ferrite (HT); Zn2p region of (g) PAN/MnZn-Ferrite, (h) PAN/MnZn Ferrite (HT).

*S.1. Stemness maintenance evaluation*

The ability of the scaffolds to support hMSC stemness was evaluated using RT-PCR. CD90 and CD105 were selected due to their roles as markers for hMSCs [1]. The results demonstrate that both doped materials maintain the cells' stemness and create a favorable environment for effective tissue regeneration.

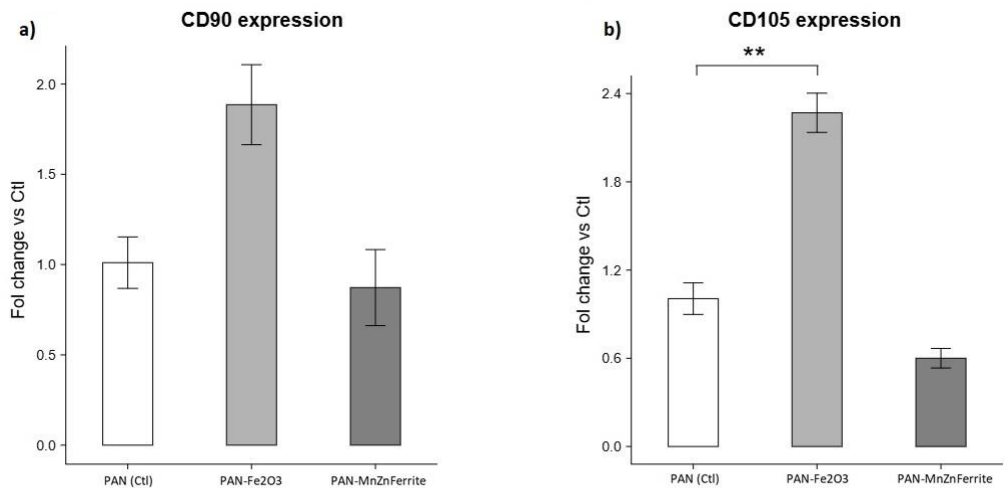

**Figure S2** – Stemness maintenance. RT-PCR results related to the expression of a)CD90 and b)CD105. **p<0.01

**Table S2** – Forward and reverse primers used to evaluate cells' adhesion, proliferation, and stemness.

| Gene | Forward primer | Reverse Primer |
|---|---|---|
| Glyceraldehyde-3-Phosphate Dehydrogenase (GAPDH)- Housekeeping | 5'- GTA TGA CAA CAG CCT CAA GAT -3' | 5'- GTC CTT CCA CGA TAC CAA AG -3' |
| Ki67 | 5'-GGA AAG TGG ACG TAG AAG AAG-3' | 5'-GCA CTG GAG TTC CCA TAA AT-3' |
| Cadherin-1 (CDH1) | 5'- CCC TTC ACA GCA GAA CTA AC -3' | 5'- TGT AGT CAC CCA CCT CTA AG -3' |
| CD90 (Thy-1) | 5'- GAC CCG TGA GAC AAA GAA G -3' | 5'- TAG TGA AGG CGG ATA AGT AGA -3' |
| CD105 (Endoglin) | 5'- CCA TCC TTG AAG TCC ATG TC -3' | 5'- GTT TAC ACT GAG GAC CAG AAG -3' |

**References**

[1]     C.S. Lin, Z.C. Xin, J. Dai, T.F. Lue, Histol Histopathol, 28, 1109-1116 (2013), http://doi.org/10.14670/hh-28.1109.